\documentclass{mems2027}

\usepackage{cite}        
\usepackage{microtype}   
\usepackage{xcolor}      
\usepackage{multirow}
\usepackage{tabularx}

\makeatletter
\def\bstctlcite{\@ifnextchar[{\@bstctlcite}{\@bstctlcite[@auxout]}}
\def\@bstctlcite[#1]#2{\@bsphack
  \@for\@citeb:=#2\do{%
    \edef\@citeb{\expandafter\@firstofone\@citeb}%
    \if@filesw\immediate\write\csname #1\endcsname{\string\citation{\@citeb}}\fi}%
  \@esphack}
\makeatother

\title{20~\NoCaseChange{GHz} Longitudinal Surface Acoustic Wave Resonators in 100-\NoCaseChange{nm} X-Cut Lithium Niobate on Silicon Carbide}
\author{Jiajun~Huang$^{*}$, Ziqian~Yao, Tzu-Hsuan~Hsu, Ian~Anderson, Joshua~Campbell, and Ruochen~Lu}
\affiliation{The University of Texas at Austin, USA}

\begin{document}
\bstctlcite{IEEEexample:BSTcontrol}
\maketitle

\begin{abstract}
Surface acoustic wave (SAW) resonators above 15~GHz require a fast acoustic mode that retains strong electromechanical coupling ($k^2$). This work reports longitudinal SAW (L-SAW) resonators designed for high $k^2$ in 100-nm X-cut lithium niobate (LN) on 4H silicon carbide (SiC), operating up to 21.7~GHz and reaching a $k^2$ of 8.1\%, a maximum Bode $Q$ of 70, and a figure of merit (FoM) of 5.7 at 20.24~GHz. In this thin-film waveguide, shrinking the wavelength ($\lambda$) keeps the normalized LN thickness ($h_\mathrm{LN}/\lambda$) large, sustaining both $k^2$ and acoustic confinement. Conversely, increasing $\lambda$ from 292 to 471~nm lowers $h_\mathrm{LN}/\lambda$ from 0.34 to 0.21 and raises the L-SAW phase velocity from 6.32 to 7.05~km/s, within 1.1\% of the 7126-m/s shear-wave velocity of 4H-SiC. As the velocity approaches this limit, more of the acoustic energy spreads into the non-piezoelectric SiC, and $k^2$ falls from its 8.1\% peak at $\lambda = 322$~nm to 1.7\% at $\lambda = 471$~nm. No L-SAW resonance is observed for $\lambda \geq 518$~nm ($h_\mathrm{LN}/\lambda \leq 0.19$), where the mode becomes leaky. Overall, the FoM stays within 5\% of its maximum for $h_\mathrm{LN}/\lambda$ between 0.26 to 0.31, defining a design window for L-SAW resonators in thin LN on SiC at K-band and beyond.
\end{abstract}


\section{Introduction}
Acoustic resonators with high electromechanical coupling ($k^2$) and quality factor ($Q$) above 10~GHz are needed for compact front-end filters in emerging 6G centimeter-wave and millimeter-wave (mmWave) bands~\cite{hagelauer2023,lu2025review}. Bulk acoustic resonators have already reached the mmWave range~\cite{yang2018ifcs,link2021ius,schaffer2022mwcl,kramer2022iedm,kramer2025apl}. Surface acoustic wave (SAW) devices are attractive because their frequency is defined lithographically, so multiple bands can be integrated on one chip, as also exploited in laterally vibrating LN resonators~\cite{gong2013tmtt,gong2014jmems}. However, the SAW frequency equals the phase velocity ($v_\mathrm{p}$) divided by the wavelength ($\lambda$), so scaling beyond 10~GHz requires either deep-sub-micrometer electrodes or a faster acoustic mode.

Thin-film lithium niobate (LN) bonded to 4H silicon carbide (SiC) combines the strong piezoelectric coupling of LN~\cite{gong2013tmtt} with a fast, low-loss substrate that confines the acoustic energy near the surface~\cite{zhang2020tmtt,du2024jmems}. On this platform, the shear-horizontal SAW (SH-SAW) offers large $k^2$ in the C-band~\cite{hsu2024jmems}, but its low $v_\mathrm{p}$ limits frequency scaling~\cite{hsu2024icmam}. The faster longitudinal SAW (L-SAW), often referred to as the longitudinal leaky SAW, is rarely used on silicon, where its velocity approaches that of the substrate and its energy leaks into the bulk; the stiffer SiC restores its confinement~\cite{hsu2025tuffc}. L-SAW resonators and filters in X-cut LN on SiC have thus been demonstrated for sub-7-GHz bands~\cite{liu2023apl,zheng2024tmtt,sun2024ims,liu2025mne,zhang2023mems} and, more recently, beyond 10~GHz~\cite{campbell2026hh,hsu2025tuffc}, while higher-order and thickness-shear modes have pushed LN-on-SiC SAW devices toward the mmWave range~\cite{campbell2024uffc,campbell2025tuffc,zhou2023eng,lee2026apl}.

Even on SiC, however, the L-SAW velocity lies close to the shear-wave velocity of 4H-SiC, $v_\mathrm{SiC}$ = 7126~m/s~\cite{kamitani1997}, so the mode is guided only above a minimum normalized LN thickness ($h_\mathrm{LN}/\lambda$). Below this limit, the L-SAW becomes leaky and its resonance degrades, as predicted theoretically~\cite{takei2024jjap} and reported for micrometer-wavelength devices in thicker films~\cite{zheng2024tmtt,sun2024ims}. This work examines the limit at sub-500-nm wavelengths in a 100-nm X-cut LN film on 4H-SiC, where varying $\lambda$ alone moves a single film from guided toward leaky operation. The resonators reach frequencies up to 21.7~GHz and reveal how the L-SAW velocity approaches the substrate limit while its coupling decays until the mode is cut off.

\begin{figure}[!t]
  \centering
  \includegraphics[width=0.92\columnwidth]{figures/fig1_device.png}
  \caption{Side view and optical image (top view, $\lambda$ = 292~nm) of the SAW resonator in 100-nm X-cut LN on 4H-SiC, with the thicknesses of each layer. The IDT is oriented $32^\circ$ from the crystal Y axis.}
  \label{fig:device}
\end{figure}

\section{Design and Simulation}
The resonator consists of aluminum (Al) interdigital transducers (IDTs) on 100-nm X-cut LN bonded to a 500-$\mu$m-thick 4H-SiC substrate (Fig.~\ref{fig:device}). The IDTs are oriented $32^\circ$ from the crystal Y axis toward Z, corresponding to Euler angles of $(90^\circ, 90^\circ, 32^\circ)$~\cite{ieee176}. In this rotated device frame, with $x_1$ along the propagation direction and $x_3$ normal to the surface, the two largest piezoelectric coefficients are $e_{11} \approx -4.6$~C/m$^2$ and $e_{34} \approx -4.5$~C/m$^2$. The coefficient $e_{11}$ couples the lateral electric field between the electrodes to the longitudinal strain $S_1$, exciting the L-SAW, whereas $e_{34}$ couples the vertical field under the electrodes to the shear strain $S_4$, exciting the SH-SAW~\cite{dai2024edl}. Their one-dimensional coupling factors peak at in-plane angles of $30.7^\circ$ and $31.7^\circ$, respectively, so this orientation efficiently excites both modes, as intended here. L-SAW-only devices have instead used $40^\circ$, which maximizes the L-SAW coupling in their thicker film~\cite{zheng2024tmtt}.

\begin{figure}[!t]
  \centering
  \includegraphics[width=0.86\columnwidth]{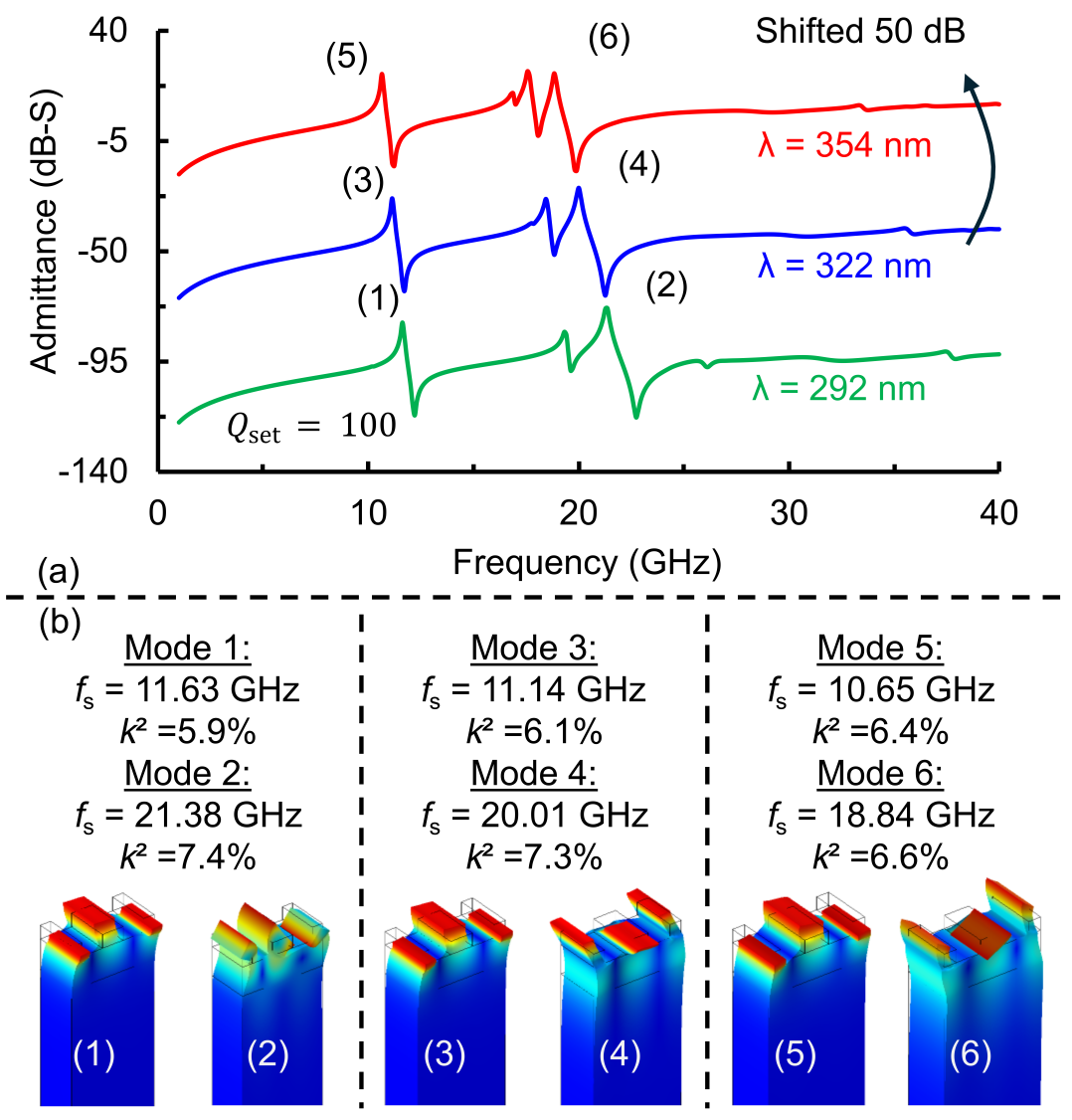}
  \caption{(a) Simulated admittance of unit-cell models with $\lambda$ = 292, 322, and 354~nm ($Q_\mathrm{set}$ = 100; curves offset by 50~dB). (b) Displacement mode shapes, series resonance frequencies ($f_\mathrm{s}$), and $k^2$ of the SH-SAW (modes 1, 3, 5) and L-SAW (modes 2, 4, 6).}
  \label{fig:sim}
\end{figure}

Figure~\ref{fig:sim} shows finite-element analysis (FEA) of periodic unit cells for $\lambda$ = 292, 322, and 354~nm in COMSOL with a uniform $Q_\mathrm{set}$ of 100. Two dominant resonances appear: an SH-SAW at 10.65--11.63~GHz with $k^2$ of 5.9--6.4\%, and an L-SAW at 18.84--21.38~GHz with $k^2$ of 6.6--7.4\%. Here, the FEA $k^2$ is $(\pi^2/8)(f_\mathrm{p}^2-f_\mathrm{s}^2)/f_\mathrm{s}^2$, which equals $(\pi^2/8)\,C_\mathrm{m}/C_0$ for an isolated resonance~\cite{lu2019jmems}.

\begin{figure}[!t]
  \centering
  \includegraphics[width=\columnwidth]{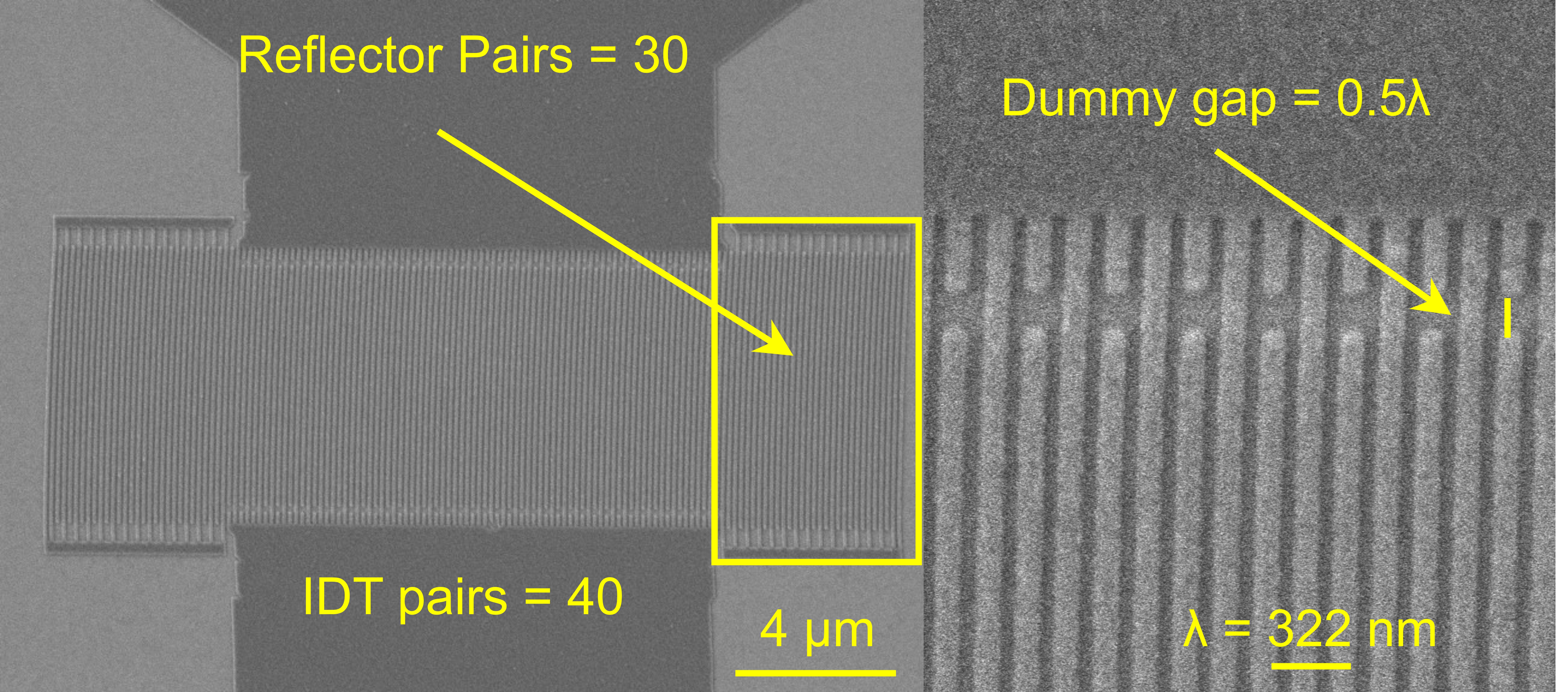}
  \caption{Scanning electron microscope (SEM) images of a fabricated resonator with $\lambda$ = 322~nm, showing the IDT, the reflectors, and the 0.5$\lambda$ dummy gap.}
  \label{fig:sem}
\end{figure}

The resonator geometry balances acoustic performance against fabrication yield (Fig.~\ref{fig:sem}). Each resonator has 40 IDT pairs with a 20$\lambda$ aperture, as in prior centimeter-band SAW resonators~\cite{hsu2024icmam}. The number of pairs and the aperture set the static capacitance and the motional resistance~\cite{morgan2007}; because the aperture scales with $\lambda$ while $f_\mathrm{s}$ scales with $1/\lambda$, the impedance of $C_0$ at resonance stays similar for all devices, while the short aperture limits both the resistance of the long, narrow fingers, which lowers $Q$, and their self-inductance, which shifts $f_\mathrm{s}$ and inflates the apparent $k^2$~\cite{campbell2024uffc}. Thirty reflector pairs per side, within the 20--60 used in prior LN-on-SiC resonators~\cite{campbell2024uffc,hsu2024jmems,lee2026apl}, should confine the energy along the propagation direction, given the strong reflectivity of the relatively thick electrodes~\cite{morgan2007}. Dummy electrodes across a 0.5$\lambda$ gap keep the metallization, and hence the velocity, nearly uniform between the aperture and the bus bars, which suppresses transverse spurious modes and scattering at the finger tips~\cite{sun2024ims}; the gap remains wide enough to be resolved without bridging. The 45-nm Al thickness follows the trade-off reported for thin-film LN SAW resonators~\cite{hsu2021jmm,hsu2024icmam}: thinner electrodes raise $f_\mathrm{s}$ but reduce $k^2$, because they cannot concentrate the strain energy near the surface, and increase the series resistance, whereas thicker, heavier electrodes lower $f_\mathrm{s}$ through mass loading and may strengthen spurious modes. Fabrication also matters: because both LN and SiC are insulating, large and dense patterns written by electron-beam lithography (EBL) accumulate charge that deflects the beam and distorts the fingers~\cite{arat2019}, so the aperture and the numbers of IDT and reflector pairs were kept moderate to preserve pattern fidelity.

\section{Fabrication and Measurement}
The devices were fabricated on a 100-nm X-cut LN on 4H-SiC wafer provided by NGK Insulators, Ltd., the platform also used in~\cite{campbell2024uffc,hsu2025tuffc}. The 45-nm-thick Al IDTs were patterned by EBL and lift-off. Resonators with $\lambda$ from 292 to 471~nm were measured in air at room temperature with an Agilent E8361C network analyzer after a full two-port calibration, with no de-embedding applied. The admittance was extracted from the measured S-parameters and fitted with a multi-branch mBVD model including series resistance and inductance~\cite{larson2000,campbell2024uffc}, and the Bode $Q$ was evaluated as $Q_\mathrm{Bode} = \omega|dS_{11}/d\omega|/(1-|S_{11}|^2)$~\cite{jin2021ius}, where $S_{11} = (1-YZ_0)/(1+YZ_0)$ is computed from the measured admittance $Y$ with $Z_0$ = 50~$\Omega$, and $Q_\mathrm{max}$ is its maximum. The measured $k^2$ is extracted from the fit as $(\pi^2/8)\,C_\mathrm{m}/C_0$ rather than from $f_\mathrm{s}$ and $f_\mathrm{p}$, which the series inductance tends to exaggerate~\cite{campbell2024uffc}, and $f_\mathrm{s}$ denotes the fitted motional series resonance.

\section{Results and Discussion}
\begin{figure}[!t]
  \centering
  \includegraphics[width=0.9\columnwidth]{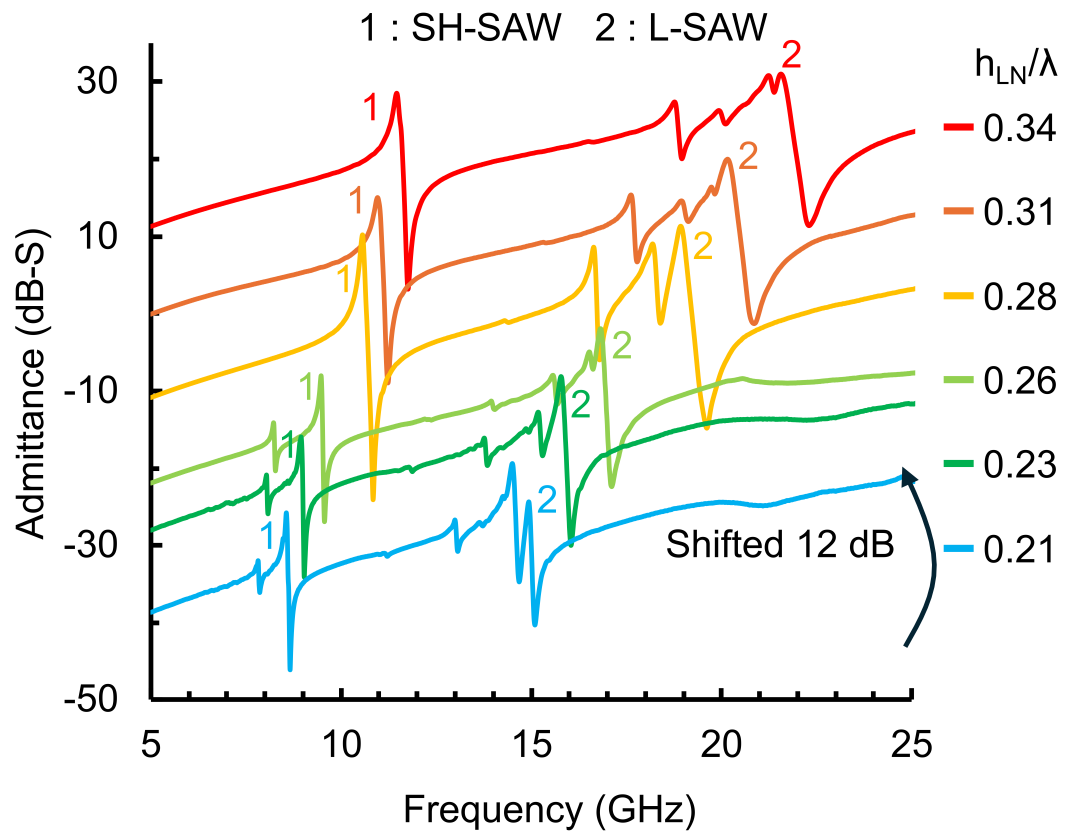}
  \caption{Measured admittance of resonators with $\lambda$ from 292 to 471~nm ($h_\mathrm{LN}/\lambda$ from 0.34 to 0.21), offset by 12~dB. Labels mark the SH-SAW (1) and L-SAW (2).}
  \label{fig:meas}
\end{figure}

Figure~\ref{fig:meas} shows the measured admittance of all six resonators, and Table~\ref{tab:summary} summarizes their performance. Beyond $\lambda$ = 322~nm, the L-SAW $k^2$ decays steadily with increasing $\lambda$ while its phase velocity ($f_\mathrm{s}\cdot\lambda$) approaches $v_\mathrm{SiC}$, to within about 1\% at $\lambda$ = 471~nm, and no L-SAW resonance is observed for $\lambda \geq 518$~nm ($h_\mathrm{LN}/\lambda \leq 0.19$). As $v_\mathrm{p}$ nears $v_\mathrm{SiC}$, the L-SAW extends deeper into the non-piezoelectric SiC until it turns leaky. This behavior agrees with the non-leaky limit of $h/\lambda \approx 0.21$ predicted for metallized X-cut LN on 4H-SiC~\cite{takei2024jjap} and with the leakage reported near the same normalized thickness in thicker films~\cite{zheng2024tmtt,sun2024ims}. Meanwhile, $Q_\mathrm{max}$ generally rises toward longer $\lambda$, so the FoM stays near its maximum over $h_\mathrm{LN}/\lambda \approx$ 0.26--0.31. This is consistent with the design windows chosen in thicker films to balance substrate leakage against spurious modes~\cite{zheng2024tmtt,sun2024ims}. The SH-SAW, being much slower than $v_\mathrm{SiC}$, stays guided over the entire range, although its $k^2$ also decreases at longer $\lambda$.

\begin{table}[!t]
  \centering
  \caption{Summary of measured resonator performance ($f_\mathrm{s}$: fitted series resonance in GHz; $k^2$ in \%; $Q$: Bode $Q_\mathrm{max}$; FoM = $k^2\cdot Q/100$).}
  \label{tab:summary}
  \setlength{\tabcolsep}{1.2pt}
  \begin{tabularx}{\columnwidth}{|c|*{8}{>{\centering\arraybackslash}X|}}
    \hline
     & \multicolumn{4}{c|}{\textbf{SH-SAW}} & \multicolumn{4}{c|}{\textbf{L-SAW}} \\ \cline{2-9}
    \textbf{$\lambda$ (nm)} & $f_\mathrm{s}$ & $k^2$ & $Q$ & FoM & $f_\mathrm{s}$ & $k^2$ & $Q$ & FoM \\ \hline
    292 & 11.50 & 5.74 & 136 & 7.81 & 21.66 & 6.70 & 61 & 4.09 \\ \hline
    322 & 10.99 & 6.80 & 143 & 9.72 & 20.24 & 8.10 & 70 & 5.67 \\ \hline
    354 & 10.57 & 7.28 & 215 & 15.65 & 19.00 & 6.15 & 92 & 5.66 \\ \hline
    389 & 9.49 & 2.08 & 212 & 4.41 & 16.87 & 3.31 & 165 & 5.46 \\ \hline
    428 & 8.97 & 1.77 & 254 & 4.50 & 15.85 & 2.78 & 155 & 4.31 \\ \hline
    471 & 8.58 & 2.55 & 274 & 6.99 & 14.96 & 1.73 & 158 & 2.73 \\ \hline
  \end{tabularx}
\end{table}


\begin{figure}[!t]
  \centering
  \includegraphics[width=0.86\columnwidth]{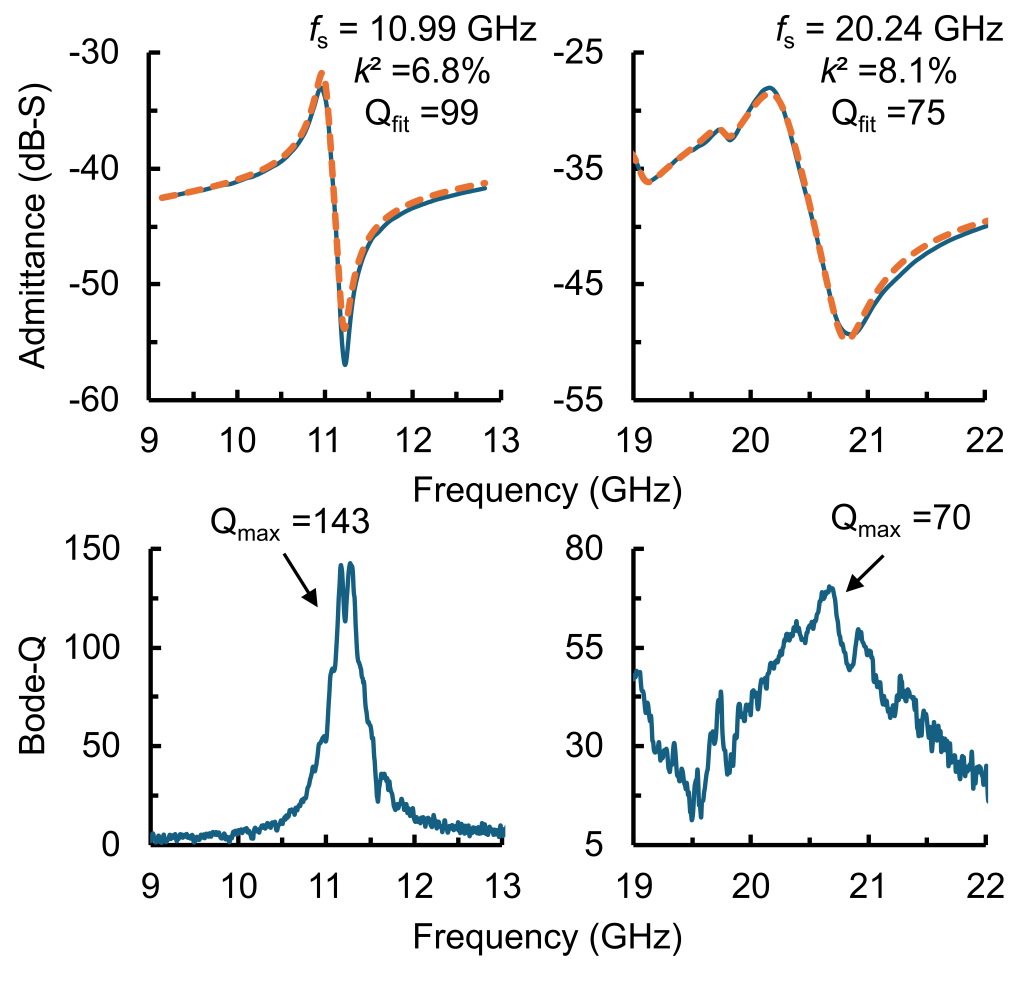}
  \caption{Measured (solid) and mBVD-fitted (dashed) admittance of the $\lambda$ = 322-nm resonator, showing the wideband response (top) and zoomed-in views of the SH-SAW and L-SAW (middle), together with the extracted Bode $Q$ around each mode (bottom). $Q_\mathrm{fit}$ is the $Q$ of the fitted mBVD motional branch.}
  \label{fig:fit}
\end{figure}

Figure~\ref{fig:fit} shows the mBVD fit of the $\lambda$ = 322-nm resonator, which provides the highest L-SAW $k^2$ and, together with the $\lambda$ = 354-nm device, the highest FoM. The measured $k^2$ exceeds the FEA values in Fig.~\ref{fig:sim} by 11--14\% for both modes at $\lambda$ = 322~nm and for the SH-SAW at 354~nm, possibly owing to fabrication deviations from the modeled geometry and to the sensitivity of the extracted $C_0$ to the parasitic branches of the model. At 292~nm, the L-SAW $k^2$ is instead below the FEA value, probably because a weak mode 0.35~GHz below it ($k^2 \approx 0.9\%$) shares its coupling; the two together reach about 7.6\%, close to the simulated 7.4\%. For both modes, the Bode $Q$ peaks near $f_\mathrm{p}$ rather than $f_\mathrm{s}$, indicating that the series resistance of the thin, narrow electrodes limits the $Q$ at resonance~\cite{chulukhadze2024hh}.

Table~\ref{tab:soa} compares these resonators with LN SAW devices above 20~GHz. Among them, the L-SAW in this work provides the highest $k^2$, up to 8.1\%, more than twice that of the 22.4-GHz L-SAW in~\cite{hsu2025tuffc}, and reaches an FoM of 5.7 at 20.24~GHz. The $Q_\mathrm{max}$ of the three devices at 19--22~GHz, below 100, is likely limited by the series resistance of the narrow 45-nm-thick fingers~\cite{chulukhadze2024hh}, which could be lowered with thicker fingers at the cost of additional mass loading.

\begin{table}[!t]
  \centering
  \caption{Comparison of LN SAW resonators above 20~GHz.}
  \label{tab:soa}
  \setlength{\tabcolsep}{1.2pt}
  \begin{tabularx}{\columnwidth}{|l|c|*{4}{>{\centering\arraybackslash}X|}}
    \hline
    \textbf{Mode [Ref.]} & \textbf{Stack} & $f_\mathrm{s}$ & $Q_\mathrm{max}$ & $k^2$ & FoM \\ \hline
    HO-SAW \cite{campbell2024uffc} & 128Y-LN/SiC & 21.4 & 70 & 3.4 & 2.38 \\ \hline
    HO-SAW \cite{zhou2023eng} & LN/SiO$_2$/SiC & 26 & 22.6 & 3.6 & 0.81 \\ \hline
    HO-SAW \cite{campbell2025tuffc} & 128Y-LN/SiC & 52 & 154 & 0.5 & 0.77 \\ \hline
    L-SAW \cite{hsu2025tuffc} & X-LN/SiC & 22.4 & 192 & 3.4 & 6.53 \\ \hline
    EG SH-SAW \cite{hsu2025tuffc} & Y-LN/SiC & 23.5 & 260 & 1.6 & 4.16 \\ \hline
    \begin{tabular}[c]{@{}l@{}}\textbf{L-SAW}\\\textbf{(this work)}\end{tabular} & \textbf{X-LN/SiC} & \textbf{20.24} & \textbf{70} & \textbf{8.1} & \textbf{5.67} \\ \hline
  \end{tabularx}
  \par\smallskip
  {\raggedright HO-SAW: higher-order SAW; EG: electrode-guided.\par}
\end{table}

\section{Conclusion}
This work demonstrated L-SAW resonators in 100-nm X-cut LN on 4H-SiC up to 21.7~GHz, with a $k^2$ of 8.1\% and an FoM of 5.7 at 20.24~GHz. As $\lambda$ increases from 292 to 471~nm ($h_\mathrm{LN}/\lambda$ from 0.34 to 0.21), the L-SAW velocity rises to within 1.1\% of the shear-wave velocity of 4H-SiC, and $k^2$ falls from its peak at $h_\mathrm{LN}/\lambda = 0.31$ to 1.7\%. No L-SAW is observed for $h_\mathrm{LN}/\lambda \leq 0.19$. This cutoff, together with the near-maximum FoM over $h_\mathrm{LN}/\lambda$ = 0.26--0.31, defines the design space for L-SAW resonators in thin LN on SiC. Lowering the electrode resistance, which likely limits $Q$, offers a path toward L-SAW filters at K-band and beyond.

\section{Acknowledgements}

This work was supported by the National Science Foundation (NSF) under CAREER Award No. 2339731 and ECCS Award No. 2530883.


\bibliographystyle{mems2027}
\bibliography{references}

\section{Contact}
*J.~Huang; jh79765@my.utexas.edu

\end{document}